\documentstyle[12pt]{article}
\def\xx{{\vphantom{-1}}}

\catcode`\@=11

\@addtoreset{equation}{section}
\def\theequation{\arabic{section}.\arabic{equation}}

\catcode`\@=11

\def\section{\@startsection{section}{1}{\z@}{3.5ex plus 1ex minus
   .2ex}{2.3ex plus .2ex}{\large\bf}}

\newskip\humongous \humongous=0pt plus 1000pt minus 1000pt

\newif\ifdtup

\def\eqnarray{\let\@currentlabel=\theequation\refstepcounter{equation}
    \global\@eqnswtrue
    \global\@eqcnt\z@\tabskip\@centering\let\\=\@eqncr
    $$\halign to \displaywidth\bgroup\@eqnsel\hskip\@centering
      $\displaystyle\tabskip\z@{##}$&\global\@eqcnt\@ne
       \hfil${{}##{}}$\hfil
      &\global\@eqcnt\tw@ $\displaystyle\tabskip\z@{##}$\hfil
       \tabskip\@centering&\llap{##}\tabskip\z@\cr}
\def\lefteqn#1{\hbox to 4\arraycolsep{$\displaystyle #1$\hss}}
\def\thesection{\arabic{section}.}

\def\appendix{\setcounter{section}{0}
        \def\thesection{Appendix.}
        \def\theequation{\Alph{section}.\arabic{equation}}}

\long\def\@makefntext#1{\parindent 0cm\noindent
\hbox to 1em{\hss$^{\@thefnmark}$}#1}
\def\IR{{\hbox{{\rm I}\kern-.2em\hbox{\rm R}}}}
\def\IH{{\hbox{{\rm I}\kern-.2em\hbox{\rm H}}}}
\def\IC{{\ \hbox{{\rm I}\kern-.6em\hbox{\bf C}}}}
\def\IZ{{\hbox{{\rm Z}\kern-.4em\hbox{\rm Z}}}}
\def\rref#1{(\ref{#1})}
\def\comp{{\scriptstyle\circ}}
\newcommand{\beq}{\begin{equation}}
\newcommand{\eeq}{\end{equation}}
\newcommand{\Tr}{\hbox{Tr}}
\newcommand{\NPB}[1]{{\sl Nucl.~Phys.}~{\bf B#1}}
\newcommand{\Ann}[1]{{\sl Ann.~Phys.}~{\bf #1}}
\newcommand{\CMP}[1]{{\sl Commun.~Math.~Phys.}~{\bf #1}}
\newcommand{\PLB}[1]{{\sl Phys.~Lett.}~{\bf B#1}}

\newcommand{\PRL}[1]{{\sl Phys.~Rev.~Lett.}~{\bf #1}}

\newcommand{\CQG}[1]{{\sl Class.~Quant.~Grav.}~{\bf #1}}
\newcommand{\PRD}[1]{{\sl Phys.~Rev.}~{\bf D#1}}

\newcommand{\JMP}[1]{{\sl J.~Math.~Phys.}~{\bf #1}}
\begin{document}
%
%
%
%
\def\citen#1{%
\edef\@tempa{\@ignspaftercomma,#1, \@end, }
\edef\@tempa{\expandafter\@ignendcommas\@tempa\@end}%
\if@filesw \immediate \write \@auxout {\string \citation {\@tempa}}\fi
\@tempcntb\m@ne \let\@h@ld\relax \let\@citea\@empty
\@for \@citeb:=\@tempa\do {\@cmpresscites}%
\@h@ld}
%
\def\@ignspaftercomma#1, {\ifx\@end#1\@empty\else
   #1,\expandafter\@ignspaftercomma\fi}
\def\@ignendcommas,#1,\@end{#1}
%
%
\def\@cmpresscites{%
 \expandafter\let \expandafter\@B@citeB \csname b@\@citeb \endcsname
 \ifx\@B@citeB\relax 
    \@h@ld\@citea\@tempcntb\m@ne{\bf ?}%
    \@warning {Citation `\@citeb ' on page \thepage \space undefined}%
 \else
    \@tempcnta\@tempcntb \advance\@tempcnta\@ne
    \setbox\z@\hbox\bgroup 
    \ifnum\z@<0\@B@citeB \relax
       \egroup \@tempcntb\@B@citeB \relax
       \else \egroup \@tempcntb\m@ne \fi
    \ifnum\@tempcnta=\@tempcntb 
       \ifx\@h@ld\relax 
          \edef \@h@ld{\@citea\@B@citeB}%
       \else 
          \edef\@h@ld{\hbox{--}\penalty\@highpenalty \@B@citeB}%
       \fi
    \else   
       \@h@ld \@citea \@B@citeB \let\@h@ld\relax
 \fi\fi%
 \let\@citea\@citepunct
}
%
\def\@citepunct{,\penalty\@highpenalty\hskip.13em plus.1em minus.1em}%
%
%
\def\@citex[#1]#2{\@cite{\citen{#2}}{#1}}%
%
%
\def\@cite#1#2{\leavevmode\unskip
  \ifnum\lastpenalty=\z@ \penalty\@highpenalty \fi 
  \ [{\multiply\@highpenalty 3 #1
      \if@tempswa,\penalty\@highpenalty\ #2\fi 
    }]\spacefactor\@m}
\let\nocitecount\relax  
%
\begin{flushright}
UCD-93-30\\
gr-qc/9309020\\
September 1993\\
\end{flushright}
\vspace{.5in}
\begin{center}
{\Large\bf
 Geometric Structures and Loop Variables\\in (2+1)-Dimensional Gravity}\\
\vspace{.4in}
{S.~C{\sc arlip}\footnote{\it email: carlip@dirac.ucdavis.edu}\\
       {\small\it Department of Physics}\\
       {\small\it University of California}\\
       {\small\it Davis, CA 95616}\\{\small\it USA}}
\end{center}

\vspace{.5in}
\addtocounter{footnote}{-1}

Ashtekar's connection representation for general relativity \cite{Ash0,Ash1}
and the closely related loop variable approach \cite{RovSmo} have
generated a good deal of excitement over the past few years.  While
it is too early to make firm predictions, there seems to be some
real hope that these new variables will allow the construction of a
consistent nonperturbative quantum theory of gravity.  Some important
progress has been made: large classes of observables have been found,
a number of quantum states have been identified, and the first steps have
been taken towards establishing a reasonable weak field perturbation
theory \cite{loops}.

Progress has been hampered, however, by the absence of a clear physical
interpretation for the observables built out of Ashtekar's new variables.
In part, the problem is simply one of unfamiliarity --- physicists
accustomed to metrics and their associated connections can find it difficult
to make the transition to densitized triads and self-dual connections.
But there is a deeper problem as well, inherent in almost any canonical
formulation of general relativity.  To define Ashtekar's variables, one must
choose a time slicing, an arbitrary splitting of spacetime into spacelike
hypersurfaces.  But real geometry and physics cannot depend on such a choice;
the true physical observables must somehow forget any details of the time
slicing, and refer only to the invariant underlying geometry.  To a certain
extent, this is already a source of trouble in classical general relativity,
where one must take care to separate physical phenomena from artifacts of
coordinate choices.  In canonical quantization, however, the problem becomes
much sharper --- all observables must be diffeomorphism invariants, and the
need to reconstruct geometry and physics from such quantities becomes
unavoidable.

In a sense, the Ashtekar program is a victim of its own success.  For the
first time, we can actually write down a large set of diffeomorphism-invariant
observables, the loop variables of Rovelli and Smolin \cite{RovSmo}.  But
although some progress has been made in defining area and volume operators
in terms of these variables \cite{area}, the goal of reconstructing spacetime
geometry from such invariant quantities remains out of reach.

The purpose of this article is to demonstrate that such a reconstruction is
possible in the simple model of (2+1)-dimensional gravity, general relativity
in two spatial dimensions plus time.  A reduction in the number of dimensions
greatly simplifies general relativity, allowing the use of powerful techniques
not readily available in the realistic (3+1)-dimensional theory.  As a
consequence, many of the specific results presented here will not readily
generalize to higher dimensions.  But the success of (2+1)-dimensional gravity
can be viewed as an ``existence proof'' for canonical quantum gravity, and
one may hope that at least some of the technical results have extensions to
our physical spacetime.

\section{(2+1)-Dimensional Gravity: From Geometry to Holonomies}

Let us begin with a brief review of (2+1)-dimensional general relativity in
first order formalism.  As our spacetime we take a three-manifold $M$,
which we shall often assume to have a topology $\IR\!\times\!\Sigma$, where
$\Sigma$ is a closed orientable surface.  The fundamental variables are a
triad $e_\mu{}^a$ --- a section of the bundle of orthonormal frames
--- and a connection on the same bundle, which can be specified by a
connection one-form $\omega_\mu{}^a{}_b$.\footnote{Indices $\mu,\,\nu\,
\rho,\dots$ are spacetime coordinate indices; $i,\,j,\,k,\dots$ are
spatial coordinate indices; and $a,\,b,\,c,\dots$ are ``Lorentz
indices,'' labeling vectors in an orthonormal basis.  Lorentz indices
are raised and lowered with the Minkowski metric $\eta_{ab}$.  This notation
is standard in papers in (2+1)-dimensional gravity, but differs from the
usual conventions for Ashtekar variables, so readers should be careful in
translation.}  The Einstein-Hilbert action can be written as
\beq
I_{\hbox{\scriptsize grav}} = \int_M\,e^a\wedge
  \left(d\omega_a + {1\over2}\epsilon_{abc}\omega^b\wedge\omega^c\right) ,
\label{a1}
\eeq
where $e^a = e_\mu{}^a dx^\mu$ and $\omega^a = {1\over2}\epsilon^{abc}
\omega_{\mu bc}dx^\mu$.  The action is invariant under local SO(2,1)
transformations,
\begin{eqnarray}
\delta e^a &=& \epsilon^{abc}e_b\tau_c \nonumber\\
\delta \omega^a &=& d\tau^a + \epsilon^{abc}\omega_b\tau_c ,
\label{a2}
\end{eqnarray}
as well as ``local translations,''
\begin{eqnarray}
\delta e^a &=& d\rho^a + \epsilon^{abc}\omega_b\rho_c \nonumber\\
\delta \omega^a &=& 0 .
\label{a3}
\end{eqnarray}
$I_{\hbox{\scriptsize grav}}$ is also invariant under diffeomorphisms of
$M$, of course, but this is not an independent symmetry: Witten has shown
\cite{Wit1} that when the triad $e_\mu{}^a$ is invertible, diffeomorphisms
in the connected component of the identity are equivalent to transformations
of the form \rref{a2}--\rref{a3}.  We therefore need only worry about
equivalence classes of diffeomorphisms that are not isotopic to the
identity, that is, elements of the mapping class group of $M$.

The equations of motion coming from the action \rref{a1} are easily
derived:
\beq
de^a + \epsilon^{abc}\omega_b\wedge e_c = 0
\label{a4}
\eeq
and
\beq
d\omega^a + {1\over2}\epsilon^{abc}\omega_b\wedge\omega_c = 0 .
\label{a5}
\eeq
These equations have four useful interpretations, which will form the basis
for our analysis:
\begin{enumerate}
\addtolength{\itemsep}{-.5ex}
\addtolength{\topsep}{-2.5ex}
\addtolength{\parindent}{1.2em}
\addtolength{\parsep}{-2.5ex}
\item We can solve \rref{a4} for $\omega$ as a function of $e$,
and rewrite \rref{a5} as an equation for $\omega[e]$.  The result is
equivalent to the ordinary vacuum Einstein field equations,
\beq
R_{\mu\nu}[g] = 0 ,
\label{a6}
\eeq
for the Lorentzian (that is, pseudo-Riemannian) metric $g_{\mu\nu} =
e_\mu{}^a e_\nu{}^b\eta_{ab}$.  In 2+1 dimensions, these field equations
are much more powerful than they are in 3+1 dimensions:  the full
Riemann curvature tensor is linearly dependent on the Ricci tensor,
\beq
R_{\mu\nu\rho\sigma} = g_{\mu\rho}R_{\nu\sigma}
+ g_{\nu\sigma}R_{\mu\rho} - g_{\nu\rho}R_{\mu\sigma}
- g_{\mu\sigma}R_{\nu\rho} - {1\over2}
(g_{\mu\rho}g_{\nu\sigma} - g_{\mu\sigma}g_{\nu\rho})R ,
\label{a7}
\eeq
so \rref{a6} actually implies that the metric $g_{\mu\nu}$ is flat. The
space of solutions of the field equations can thus be identified with the
set of flat Lorentzian metrics on $M$.
\item As a second alternative, note that equation \rref{a5} depends only on
$\omega$ and not $e$.  In fact, \rref{a5} is simply the requirement that
the curvature of $\omega$ vanish, that is, that $\omega$ be a flat SO(2,1)
connection.  Moreover, equation \rref{a4} can be interpreted as the statement
that $e$ is a cotangent vector to the space of flat SO(2,1) connections;
indeed, if $\omega(s)$ is a curve in the space of flat connections, the
derivative of \rref{a5} gives
\beq
d\left({d\omega^a\over ds}\right) +
 \epsilon^{abc}\omega_b\wedge\left({d\omega_c\over ds}\right) = 0 ,
\label{a8}
\eeq
which can be identified with \rref{a4} with
\beq
e^a = {d\omega^a\over ds} .
\label{a9}
\eeq

To determine the physically inequivalent solutions of the field equations,
we must still factor out the gauge transformations \rref{a2}--\rref{a3}.
The local Lorentz transformations \rref{a2} act on $\omega$ as ordinary
SO(2,1) gauge transformations, and tell us that only gauge equivalence
classes of flat SO(2,1) connections are relevant.  Let us denote the space
of such equivalence classes as $\tilde{\cal N}$.  The transformations of
$e$ can once again be interpreted as statements about the cotangent space:
if we consider a curve $\tau(s)$ of SO(2,1) transformations, it is easy
to check that the first equations of \rref{a2} and \rref{a3} follow
from differentiating the second equation of \rref{a2}, with
\beq
\rho^a = {d\tau^a\over ds}
\label{a10}
\eeq
and $e^a$ as in \rref{a9}.  A solution of the field equations is
thus determined by a point in the cotangent bundle $T^*\tilde{\cal N}$.

Now, a flat connection on the frame bundle of $M$ is determined by its
holonomies, that is, by a homomorphism
\beq
\rho: \pi_1(M)\rightarrow \hbox{SO(2,1)} ,
\label{a11}
\eeq
and gauge transformations act on $\rho$ by conjugation.  We can therefore
write
\begin{eqnarray}
\tilde{\cal N} &=& \hbox{Hom}(\pi_1(M),\hbox{SO(2,1)})/\sim ,\nonumber\\
\rho_1&\sim&\rho_2 \ \ \hbox{if}\ \ \rho_2 = h\cdot\rho_1\cdot h^{-1},
\quad h\in\hbox{SO(2,1)} .
\label{a12}
\end{eqnarray}
It remains for us to factor out the diffeomorphisms that are not in the
component of the identity, the mapping class group.  These transformations
act on $\tilde{\cal N}$ through their action as a group of automorphisms
of $\pi_1(M)$, and in many interesting cases --- for example, when $M$ has
the topology $\IR\!\times\!\Sigma$ --- this action comprises the entire
set of outer automorphisms of $\pi_1(M)$ \cite{Harvey}.  If we denote
equivalence under this action by ${\sim}^\prime$, and let
$\tilde{\cal N}/{\sim}^\prime = \cal N$,
we can express the space of solutions of the field equations
\rref{a4}--\rref{a5} as\footnote{Strictly speaking, one more subtlety
remains.  The space of homomorphisms \rref{a12} is not always connected,
and it is often the case that only one connected component corresponds
to physically admissible spacetimes.  See \cite{Gold,Mess} for the
mathematical structure and \cite{Wit1,Mess,Carmeas,LuoMar} for physical
implications.} $T^*{\cal N}$.

When $M$ has the topology $\IR\!\times\!\Sigma$, this description can be
further refined.  In that case, the space $\tilde{\cal N}$ --- or at
least the physically relevant connected component of $\tilde{\cal N}$
--- is homeomorphic to the Teichm\"uller space of $\Sigma$, and $\cal N$ is
the corresponding moduli space \cite{Gold,Gold2}.  The set of vacuum
spacetimes can thus be identified with the cotangent bundle of the moduli
space of $\Sigma$, and many powerful results from Riemann surface theory
become applicable.
\item A third approach is available if $M$ has the topology
$\IR\!\times\!\Sigma$.  For such a topology, it is useful to split
the field equations into spatial and temporal components.  Let us write
\begin{eqnarray}
d &=& \tilde d + dt\,\partial_0, \nonumber\\
e^a &=& \tilde e^a + e_0{}^a dt, \\
\omega^a &=& \tilde \omega^a + \omega_0{}^a dt .\nonumber
\label{a13}
\end{eqnarray}
(This decomposition can be made in a less explicitly coordinate-dependent
manner --- see, for example, \cite{Smolin} --- but the final results are
unchanged.)  The spatial projections of \rref{a4}--\rref{a5} take the same
form as the original equations, with all quantities replaced by their
``tilded'' spatial equivalents.  As above, solutions may therefore be
labeled by classes of homomorphisms, now from $\pi_1(\Sigma)$ to SO(2,1),
and the corresponding cotangent vectors.  The temporal components of the
field equations, on the other hand, now become
\begin{eqnarray}
\partial_0\, \tilde e^a &=& \tilde d e_0{}^a
+ \epsilon^{abc}\tilde\omega_b e_{0c} + \epsilon^{abc}\tilde e_b \omega_{0c}
\nonumber\\
\partial_0\, \tilde\omega^a &=& \tilde d \omega_0{}^a
+ \epsilon^{abc}\tilde\omega_b \omega_{0c} .
\label{a14}
\end{eqnarray}
Comparing to \rref{a2}--\rref{a3}, we see that the time development of
$(\tilde e,\tilde\omega)$ is entirely described by a gauge transformation,
with  $\tau^a = \omega_0{}^a$ and $\rho^a = e_0{}^a$.

This is consistent with our previous results, of course.  For a topology
of the form $\IR\!\times\!\Sigma$, the fundamental group is simply that
of $\Sigma$, and an invariant description in terms of holonomies should
not be able to detect a particular choice of spacelike slice.  Equation
\rref{a14} shows in detail how this occurs: motion in coordinate time is
merely a gauge transformation, and is therefore invisible to the holonomies.
But the central dilemma described in the introduction now stands out sharply.
For despite equation \rref{a14}, solutions of the (2+1)-dimensional field
equations are certainly not static {\em as geometries} --- they do not,
in general, admit timelike Killing vectors.  The real physical dynamics
has somehow been hidden by this analysis, and must be uncovered if we
are to find a sensible physical interpretation of our solutions.  This
puzzle is an example of the notorious ``problem of time'' in gravity
\cite{Kuchar}, and exemplifies one of the basic issues that must be
resolved in order to construct a sensible quantum theory.
\item A final approach to the field equations \rref{a4}--\rref{a5} was
suggested by Witten \cite{Wit1}, who observed that the triad $e$ and the
connection $\omega$ could be combined to form a single connection on an
ISO(2,1) bundle.  ISO(2,1), the three-dimensional Poincar\'e group, has a
Lie algebra with generators ${\cal J}^a$ and ${\cal P}^b$ and commutation
relations
\beq
\left[{\cal J}^a, {\cal J}^b\right] = \epsilon^{abc}{\cal J}_c ,\qquad
\left[{\cal J}^a, {\cal P}^b\right] = \epsilon^{abc}{\cal P}_c ,\qquad
\left[{\cal P}^a, {\cal P}^b\right] = 0 .
\label{a15}
\eeq
If we write a single connection one-form
\beq
A = e^a{\cal P}_a + \omega^a{\cal J}_a
\label{a16}
\eeq
and define a ``trace,'' an invariant inner product on the Lie algebra, by
\beq
\Tr\left({\cal J}^a{\cal P}^b\right) = \eta^{ab} , \qquad
Tr\left({\cal J}^a{\cal J}^b\right) = Tr\left({\cal P}^a{\cal P}^b\right)
= 0 ,
\label{a17}
\eeq
then it is easy to verify that the action \rref{a1} is simply the
Chern-Simons action \cite{WitJones} for $A$,
\beq
I_{\hbox{\scriptsize CS}} = {1\over2}\int_M\,\Tr
  \left\{ A\wedge dA + {2\over3}A\wedge A\wedge A\right\} .
\label{a18}
\eeq

The field equations now reduce to the requirement that $A$ be a flat ISO(2,1)
connection, and the gauge transformations \rref{a2}--\rref{a3} can be
identified with standard ISO(2,1) gauge transformations.  Imitating the
arguments of our second interpretation, we should therefore expect solutions
of the field equations to be characterized by gauge equivalence classes of
flat ISO(2,1) connections, that is, by homomorphisms in the space
\begin{eqnarray}
\tilde{\cal M} &=& \hbox{Hom}(\pi_1(M),\hbox{ISO(2,1)})/\sim ,\nonumber\\
\rho_1&\sim&\rho_2 \ \ \hbox{if}\ \ \rho_2 = h\cdot\rho_1\cdot h^{-1},
\quad h\in\hbox{ISO(2,1)} .
\label{a19}
\end{eqnarray}

To relate this description to our previous results, note that ISO(2,1) is
itself a cotangent bundle with base space SO(2,1).  Indeed, a cotangent
vector at the point $\Lambda_1\!\in\!\hbox{SO(2,1)}$ can be written in the
form $d\Lambda_1^\xx \Lambda_1^{\,-1}$, and the multiplication law
\begin{eqnarray}
(\Lambda_1^\xx,\,d\Lambda_1^\xx \Lambda_1^{\,-1})\,\cdot\,
  (\Lambda_2^\xx,\,d\Lambda_2^\xx \Lambda_1^{\,-1}) &=&
  (\Lambda_1^\xx\Lambda_2^\xx,\,d(\Lambda_1^\xx\Lambda_2^\xx)
  (\Lambda_1^\xx\Lambda_2^\xx)^{\,-1})\nonumber\\
  &=& (\Lambda_1^\xx\Lambda_2^\xx,
  d\Lambda_1^\xx \Lambda_1^{\,-1} +
  \Lambda_1^\xx(d\Lambda_2^\xx \Lambda_2^{\,-1})\Lambda_1^{\,-1})
\label{a20}
\end{eqnarray}
may be recognized as the standard semidirect product composition law
for Poincar\'e transformations.  The space of homomorphisms from $\pi_1(M)$
to ISO(2,1) inherits this cotangent bundle structure in an obvious way,
leading to the identification $\tilde{\cal M} \approx T^*\tilde{\cal N}$,
where $\tilde{\cal N}$ is the space of homomorphisms \rref{a12}.  It
remains for us to factor out the mapping class group.  But this group
acts in \rref{a12} and \rref{a19} as the same group of automorphisms of
$\pi_1(M)$;  writing the quotient as $\tilde{\cal M}/{\sim}' = {\cal M}$,
we thus see that ${\cal M}\approx T^*{\cal N}$.
\end{enumerate}

Of these four approaches to the (2+1)-dimensional field equations, only
the first corresponds directly to our usual picture of spacetime physics.
Trajectories of physical particles, for instance, are geodesics in the
flat manifolds of this description.  The second approach, on the other
hand, is the one that is closest to the loop variable picture in
(3+1)-dimensional gravity.  The loop variables of Rovelli and Smolin
\cite{Ash1,RovSmo,RovSmo2,JacSmo} may be expressed as follows.  Let
\beq
U[\gamma,x] = P\exp\left\{\int_\gamma \omega^a{\cal J}_a \right\}
\label{a21}
\eeq
be the holonomy of the connection one-form $\omega^a$ around a closed
path $\gamma(t)$ based at $\gamma(0)=x$.  (Here, $P$ denotes path
ordering, and the basepoint $x$ specifies the point at which the path
ordering begins.)  We then define
\beq
{\cal T}^0[\gamma] = \Tr\, U[\gamma,x]
\label{a22}
\eeq
and
\beq
{\cal T}^1[\gamma] =  \int_\gamma dt\, \Tr\left\{
  U[\gamma, x(t)]\, e_\mu{}^a(\gamma(t)){dx^\mu\over dt}(\gamma(t))
  {\cal J}_a\right\} .
\label{a23}
\eeq
 ${\cal T}^0[\gamma]$ is thus the trace of the SO(2,1) holonomy around
$\gamma$, while ${\cal T}^1[\gamma]$ is essentially a cotangent vector to
${\cal T}^0[\gamma]$: indeed, given a curve $\omega(s)$ in the space of
flat SO(2,1) connections, we can differentiate \rref{a22} to obtain
\beq
{d\ \over ds} T^0[\gamma] = \int_\gamma \Tr\left( U[\gamma,x(t)]\,
  {d\omega^a\over ds}(\gamma(t)){\cal J}_a\right) ,
\label{a24}
\eeq
and we have already seen that the derivative $d\omega^a/ds$ can be
identified with the triad $e^a$.

In 3+1 dimensions, the variables ${\cal T}^0$ and ${\cal T}^1$ depend on
particular loops $\gamma$, and considerable work is still needed to
construct diffeomorphism-invariant observables that depend only on knot
classes.  In 2+1 dimensions, on the other hand, the loop variables are
already invariant, at least under diffeomorphisms isotopic to the identity.
The key difference is that in 2+1 dimensions the connection $\omega$ is
flat, so the holonomy $U[\gamma,x]$ depends only on the homotopy class of
$\gamma$.  Some care must be taken in handling the mapping class group,
which acts nontrivially on ${\cal T}^0$ and ${\cal T}^1$; this issue has
been investigated in a slightly different context by Nelson and Regge
\cite{NR1}.

Of course, ${\cal T}^0$ and ${\cal T}^1$ are not quite the equivalence
classes of holonomies of our interpretation number two above:
${\cal T}^0[\gamma]$ is not a holonomy, but only the {\em trace} of a
holonomy.  But knowledge of ${\cal T}^0[\gamma]$ for a large enough
set of homotopically inequivalent curves may be used to reconstruct a
point in the space $\tilde{\cal N}$ of equation \rref{a12}, and indeed,
the loop variables can serve as local coordinates on $\tilde{\cal N}$
\cite{holon,holon2,NR3}.

\section{Geometric Structures: From Holonomies to Geometry}

The central problem described in the introduction can now be made explicit.
Spacetimes in 2+1 dimensions can be characterized \`a la Ashtekar, Rovelli,
and Smolin as points in the cotangent bundle $T^*{\cal N}$, our
description number two of the last section.  Such a description is fully
diffeomorphism invariant, and provides a natural starting point for
quantization.  But our intuitive geometric picture of a (2+1)-dimensional
spacetime is that of a manifold $M$ with a flat metric --- description
number one --- and only in this representation do we know how to connect
the mathematics with ordinary physics.  Our goal is therefore to provide
a translation between these two descriptions.

To proceed, let us investigate the space of flat spacetimes in a bit
more detail.  If $M$ is topologically trivial, the vanishing of the
curvature tensor implies that $(M,g)$ is simply ordinary Minkowski space
$(V^{2,1},\eta)$, or at least to some subset of $(V^{2,1},\eta)$ that
can be extended to the whole of Minkowski space.  If the spacetime
topology is nontrivial, $M$ can still be covered by contractible coordinate
patches $U_i$ that are each isometric to $V^{2,1}$, with the standard
Minkowski metric $\eta_{\mu\nu}$ on each patch.  The geometry is then
encoded in the transition functions $\gamma_{ij}$ on the intersections
$U_i\cap U_j$, which determine how these patches are glued together.
Moreover, since the metrics in $U_i$ and $U_j$ are identical, these
transition functions must be isometries of $\eta_{\mu\nu}$, that is,
elements of the Poincar\'e group ISO(2,1).

Such a construction is an example of what Thurston calls a geometric
structure \cite{Thur,Canary,Gold3,SullThur}, in this case a Lorentzian
or (ISO(2,1),$V^{2,1}$) structure.  In general, a $(G,X)$ manifold is
one that is locally modeled on $X$, just as an ordinary $n$-dimensional
manifold is modeled on $\IR^n$.  More precisely, let $G$ be a Lie group
that acts analytically on some $n$-manifold $X$, the model space, and
let $M$ be another $n$-manifold.  A $(G,X)$ structure on $M$ is then
a set of coordinate patches $U_i$ covering $M$ with ``coordinates''
$\phi_i: U_i\rightarrow X$ taking their values in the model space
and with transition functions $\gamma_{ij}^\xx =
\phi_i^\xx\comp\phi_j^{\ -1}|U_i\cap U_j$ in $G$.  While this
general formulation may not be widely known, specific examples
are familiar: for example, the uniformization theorem for Riemann
surfaces implies that any surface of genus $g>1$ admits an $(\IH^2,
\hbox{PSL(2,$\IR$)})$ structure.

A fundamental ingredient in the description of a $(G,X)$ structure is
its holonomy group, which can be viewed as a measure of the failure of a
single coordinate patch to extend around a closed curve.  Let $M$ be a
$(G,X)$ manifold containing a closed path $\gamma$.  We can cover
$\gamma$ with coordinate charts
\beq
\phi_i: U_i\rightarrow X,\qquad i=1,\dots,n
\label{b1}
\eeq
with constant transition functions $g_i\in G$ between $U_i$ and $U_{i+1}$,
i.e.,
\begin{eqnarray}
\phi_i|U_i\cap U_{i+1} &=& g_i\comp \phi_{i+1}|U_i\cap U_{i+1}\nonumber\\
\phi_n|U_n\cap U_{1} &=& g_n\comp \phi_{1}|U_n\cap U_{1} .
\label{b2}
\end{eqnarray}
Let us now try to analytically continue the coordinate $\phi_1$ from the
patch $U_1$ to the whole of $\gamma$.  We can begin with a coordinate
transformation in $U_2$ that replaces $\phi_2$ by ${\phi_2}'=g_1\comp\phi_2$,
thus extending $\phi_1$ to $U_1\cup U_2$.  Continuing this process along
the curve, with ${\phi_j}' = g_1\comp\dots\comp g_{j-1}\comp\phi_j$, we will
eventually reach the final patch $U_n$, which again overlaps $U_1$.  If the
new coordinate function ${\phi_n}'=g_1\comp\dots\comp g_{n-1}\comp\phi_n$
happens to agree with $\phi_1$ on $U_n\cap U_1$, we will have succeeded in
covering $\gamma$ with a single patch.  Otherwise, the holonomy $H$, defined
as $H(\gamma) = g_1\comp\dots\comp g_n$, measures the obstruction to such a
covering.

It may be shown that the holonomy of a curve $\gamma$ depends only on its
homotopy class \cite{Thur}.  In fact, the holonomy defines a homomorphism
\beq
H: \pi_1(M)\rightarrow G .
\label{b3}
\eeq
Note that if we pass from $M$ to its universal covering space $\widetilde
M$, we will no longer have noncontractible closed paths, and $\phi_1$ will
be extendable to all of $\widetilde M$.  The resulting map $D\!:\!\widetilde
M\!\rightarrow\!X$ is called the developing map of the $(G,X)$ structure.
At least in simple examples, $D$ embodies the classical geometric
picture of development as ``unrolling'' --- for instance, the unwrapping
of a cylinder into an infinite strip.

The homomorphism $H$ is not quite uniquely determined by the geometric
structure, since we are free to act on the model space $X$ by a fixed
element $h\in G$, thus changing the transition functions $g_i$ without
altering the $(G,X)$ structure of $M$.  It is easy to see that such a
transformation has the effect of conjugating $H$ by $h$, and it is not
hard to prove that $H$ is in fact unique up to such conjugation \cite{Thur}.
For the case of a Lorentzian structure, where $G = \hbox{ISO(2,1)}$, we
are thus led to a space of holonomies of precisely the form \rref{a19}.

This identification is not a coincidence.  Given a $(G,X)$ structure on a
manifold $M$, it is straightforward to define a corresponding flat $G$
bundle \cite{Gold3}.  To do so, we simply form the product $G\times U_i$
in each patch --- giving the local structure of a $G$ bundle --- and use
the transition functions $\gamma_{ij}$ of the geometric structure to glue
together the fibers on the overlaps.  It is then easy to verify that the
flat connection on the resulting bundle has a holonomy group isomorphic
to the holonomy group of the geometric structure.

We can now try to reverse this process, and use one the holonomy groups of
equation \rref{a19} --- approach number three to the field equations ---
to define a Lorentzian structure on $M$, reproducing approach number one.
In general, this step may fail: the holonomy group of a $(G,X)$ structure
is not necessarily sufficient to determine the full geometry.  For
spacetimes, it is easy to see what can go wrong.  If we start with a
flat three-manifold $M$ and simply cut out a ball, we can obtain a new
flat manifold without affecting the holonomy of the geometric structure.
This is a rather trivial change, however, and we would like to show that
nothing worse can go wrong.

Mess \cite{Mess} has investigated this question for the case of spacetimes
with topologies  of the form $\IR\!\times\!\Sigma$.  He shows that the
holonomy group determines a unique ``maximal'' spacetime $M$ ---
specifically, a  spacetime constructed as a domain of dependence of a
spacelike surface $\Sigma$.  Mess also demonstrates that the holonomy
group $H$ acts properly discontinuously on a region $W\!\subset\!V^{2,1}$
of Minkowski space, and that $M$ can be obtained as the quotient space
$W/H$.  This quotient construction can be a powerful tool for obtaining
a description of $M$ in reasonably standard coordinates, for instance
in a time slicing by surfaces of constant mean curvature.

For topologies more complicated than $\IR\!\times\!\Sigma$, I know of
very few general results.  But again, a theorem of Mess is relevant:
if $M$ is a compact three-manifold with a flat, nondegenerate,
time-orientable Lorentzian metric and a strictly spacelike boundary,
then $M$ necessarily has the topology $\IR\!\times\!\Sigma$, where
$\Sigma$ is a closed surface homeomorphic to one of the boundary
components of $M$.  This means that for spatially closed three-dimensional
universes, topology change is classically forbidden, and the full topology
is uniquely fixed by that of an initial spacelike slice.  Hence, although
more exotic topologies may occur in some approaches to quantum gravity,
it is not physically unreasonable to restrict our attention to spacetimes
$\IR\!\times\!\Sigma$.

To summarize, we now have a procedure --- valid at least for spacetimes
of the form $\IR\!\times\!\Sigma$ --- for obtaining a flat geometry
from the invariant data given by Ashtekar-Rovelli-Smolin loop variables.
First, we use the loop variables determine a point in the cotangent bundle
$T^*{\cal N}$, establishing a connection to our second approach to the
field equations.  Next, we associate that point with an ISO(2,1) holonomy
group $H\!\in\!{\cal M}$, as in our approach number four.  Finally, we
identify the group $H$ with the holonomy group of a Lorentzian structure
on $M$, thus determining a flat spacetime of approach number one.  In
particular, if we can solve the (difficult) technical problem of finding
an appropriate fundamental region $W\!\subset\!V^{2,1}$ for the action of
$H$, we can write $M$ as a quotient space $W/H$.

This procedure has been investigated in detail for the case of a torus
universe, $\IR\!\times\!T^2$, in references \cite{Carobs} and \cite{LuoMar}.
For a universe containing point particles, it is implicit in the early
descriptions of Deser et al.\ \cite{DJtH}, and is explored in some
detail in \cite{Carmeas}.  For the (2+1)-dimensional black hole, the
geometric structure can be read off from references \cite{HenTeit} and
\cite{Mann}.\footnote{For the black hole, a cosmological constant must
be added to the field equations.  Instead of being flat, the resulting
spacetime has constant negative curvature, and the geometric structure
becomes an $(\IH^{2,1},\hbox{SO(2,2)})$ structure.  A related result for
the torus will appear in \cite{CarNel}.}   And although it is never stated
explicitly,
the recent work of 't Hooft \cite{tH} and Waelbroeck \cite{Wael} is really
a description of flat spacetimes in terms of Lorentzian structures.

\section{Quantization and Geometrical Observables}

Our discussion so far has been strictly classical.  I would like to conclude
by briefly describing some of the issues that arise if we attempt to
quantize (2+1)-dimensional gravity.

The canonical quantization of a classical system is by no means uniquely
defined, but most approaches have some basic features in common.  A
classical system is characterized by its phase space, a $2N$-dimensional
symplectic manifold $\Gamma$, with local coordinates consisting of $N$
position variables and $N$ conjugate momenta.  Classical observables are
functions of the positions and momenta, that is, maps $f,g,\dots$ from
$\Gamma$ to $\IR$.  The symplectic form $\Omega$ on $\Gamma$ determines a
set of Poisson brackets $\{f,g\}$ among observables, and hence induces a
Lie algebra structure on the space of observables.  To quantize such a
system, we are instructed to replace the classical observables with
operators and the Poisson brackets with commutators; that is, we are to
look for an irreducible representation of this Lie algebra as an algebra
of operators acting on some (normally $N$-dimensional) Hilbert space.

As stated, this program cannot be carried out: Van Hove showed in 1951
that in general, no such irreducible representation of the full Poisson
algebra of classical observables exists \cite{vanH}.  In practice, we must
therefore choose a subalgebra of ``preferred'' observables to quantize,
one that must be small enough to permit a consistent representation and
yet big enough to generate a large class of classical observables
\cite{Isham}.  Ordinarily, the resulting quantum theory will depend on
this choice of preferred observables, and we will have to look hard for
physical and mathematical justifications for our selection.

In simple classical systems, there is often an obvious set of preferred
observables --- the positions and momenta of point particles, for
instance, or the fields and their canonical momenta in a free field theory.
For gravity, on the other hand, such a natural choice seems difficult to
find.  In 2+1 dimensions, where a number of approaches to quantization
can be carried out explicitly, it is known that different choices of
variables lead to genuinely different quantum theories
\cite{CarDirac,Carsix,Marolf}.

In particular, each of the four interpretations of the field equations
discussed above suggests its own set of fundamental observables.  In the
first interpretation --- solutions as flat spacetimes --- the natural
candidates are the metric and its canonical momentum on some spacelike
surface.  But these quantities are not diffeomorphism invariant, and it
seems that the best we can do is to define a quantum theory in some
particular, fixed time slicing \cite{Carobs,Moncrief,HosNak}.  This is a
rather undesirable situation, however, since the choice of such a slicing
is arbitrary, and there is no reason to expect the quantum theories coming
from different slicings to be equivalent.

In our second interpretation --- solutions as classes of flat connections
and their cotangents --- the natural observables are points in the bundle
$T^*{\cal N}$.  These are diffeomorphism invariant, and quantization is
relatively straightforward; in particular, the appropriate symplectic
structure for quantization is just the natural symplectic structure of
$T^*{\cal N}$ as a cotangent bundle.  The procedure for quantizing such a
cotangent bundle is well-established \cite{Ash1}, and there seem to be no
fundamental difficulties in constructing the quantum theory.  But now, just
as in the classical theory, the physical interpretation of the quantum
observables is obscure.

It is therefore natural to ask whether we can extend the classical
relationships between these approaches to the quantum theories.  At least
for simple topologies, the answer is positive.  The basic strategy is as
follows.

We begin by choosing a set of physically interesting classical observables
of flat spacetimes.  For example, it is often possible to uniquely foliate
a spacetime by spacelike hypersurfaces of constant mean extrinsic curvature
$\Tr K$; the intrinsic and extrinsic geometries of such slices are useful
observables with clear physical interpretations.  Let us denote these
variables generically as $Q(T)$, where the parameter $T$ labels the
time slice on which the $Q$ are defined (for instance, $T=\Tr K$).

Classically, such observables can be determined --- at least in principle
--- as functions of the geometric structure, and thus of the ISO(2,1)
holonomies $\rho$,
\beq
Q = Q[\rho,T] .
\label{c1}
\eeq
We now adopt these holonomies as our preferred observables for quantization,
obtaining a Hilbert space $L^2({\cal N})$ and a set of operators $\hat\rho$.
Finally, we translate \rref{c1} into an operator equation,
\beq
\widehat Q = \widehat Q[\hat\rho,T] ,
\label{c2}
\eeq
thus obtaining a set of diffeomorphism-invariant but ``time-dependent''
quantum observables to represent the variables $Q$.  Some ambiguity
will remain, since the operator ordering in \rref{c2} is rarely unique,
but in the examples studied so far, the requirement of mapping class group
invariance seems to place major restrictions on the possible orderings
\cite{CarDirac}.

This program has been investigated in some detail for the simplest
nontrivial topology, $M=\IR\!\times\!T^2$ \cite{Carobs,CarDirac}.
There, a natural set of ``geometric'' variables are the modulus $\tau$
of a toroidal slice of constant mean curvature $\Tr K\!=\!T$ and its
conjugate momentum $p_\tau$.  These can be expressed explicitly in
terms of a set of loop variables that characterize the ISO(2,1)
holonomies of the spacetime.  Following the program outlined above,
one obtains a one-parameter family of diffeomorphism-invariant operators
$\hat\tau(T)$ and $\hat p_\tau(T)$ that describe the physical evolution
of a spacelike slice.  The $T$-dependence of these operators can be
described by a set of Heisenberg equations of motion,
\beq
{d\hat\tau\over dT} = i\left[ \hat H, \hat\tau \right], \qquad
{d\hat p_\tau\over dT} = i\left[ \hat H, \hat p_\tau \right],
\label{c3}
\eeq
with a Hamiltonian $\hat H[\hat\tau,\hat p_\tau, T]$ that can again be
calculated explicitly.

Let me stress that despite the familiar appearance of \rref{c3}, the
parameter $T$ is {\em not} a time coordinate in the ordinary sense; the
operators $\hat\tau(T)$ and $\hat p_\tau(T)$ are fully diffeomorphism
invariant.  We thus have a kind of ``time dependence without time
dependence,'' an expression of dynamics in terms of operators that are
individually constants of motion.  For more complicated topologies, such
an explicit construction seems quite difficult, although Unruh and Newbury
have taken some interesting steps in that direction \cite{Unruh}.
Ideally, one would like to find some kind of perturbation theory for
geometrical variables like $\widehat Q$, but little progress has yet
been made in this direction.

The specific constructions I have described here are unique to 2+1
dimensions, of course.  But I believe that some of the basic features are
likely to extend to realistic (3+1)-dimensional gravity.  The quantization
of holonomies is a form of ``covariant canonical quantization''
\cite{Ashccq,Crn}, or quantization of the space of classical solutions.
We do not yet understand the classical solutions of the
(3+1)-dimensional field equations well enough to duplicate such a
strategy, but a similar approach may be useful in minisuperspace models.
The invariant but $T$-dependent operators $\hat\tau(T)$ and $\hat p_\tau(T)$
are examples of Rovelli's ``evolving constants of motion'' \cite{Rovelli},
whose use has also been suggested in (3+1)-dimensional gravity.  Finally,
our simple model has strikingly confirmed the power of the Rovelli-Smolin
loop variables.  A full extension to 3+1 dimensions undoubtedly remains a
distant goal, but for the first time in years, there seems to be some real
cause for optimism.

\vspace{.1cm}
\begin{flushleft}
\large\bf Acknowledgements
\end{flushleft}
This work was supported in part by U.S.\ Department of Energy grant
DE-FG03-91ER40674.

\end{document}